\documentstyle[graphicx,12pt]{article}
\begin{document}

\small
\hoffset=-1truecm
\voffset=-2truecm
\title{\bf The Casimir effect for parallel plates in the spacetime
with a fractal extra compactified dimension}
\author{Hongbo Cheng\footnote {E-mail address:
hbcheng@sh163.net}\\
Department of Physics, East China University of Science and
Technology,\\ Shanghai 200237, China\\
The Shanghai Key Laboratory of Astrophysics,\\ Shanghai 200234,
China}

\date{}
\maketitle

\begin{abstract}
The Casimir effect for massless scalar fields satisfying Dirichlet
boundary conditions on the parallel plates in the presence of one
fractal extra compactified dimension is analyzed. We obtain the
Casimir energy density by means of the regularization of multiple
zeta function with one arbitrary exponent. We find a limit on the
scale dimension like $\delta>\frac{1}{2}$ to keep the negative
sign of the renormalized Casimir energy which is the difference
between the regularized energy for two parallel plates and the one
with no plates. We derive and calculate the Casimir force relating
to the influence from the fractal additional compactified
dimension between the parallel plates. The larger scale dimension
leads to the greater revision on the original Casimir force. The
two kinds of curves of Casimir force in the case of
integer-numbered extra compactified dimension or fractal one are
not superposition, which means that the Casimir force show whether
the dimensionality of additional compactified space is integer or
fraction.
\end{abstract}
\vspace{4cm} \hspace{1cm} PACS number(s): 04.50.+h, 11.10Kk,
04.62.+v

\newpage

During the investigation of quantum gravity, more attentions have
been paid to the fractal universe [1]. It may be better to
describe the spacetime whose scale is on the Planck order in
virtue of fractal geometry with some non-integer dimensions.
Within this kind of background some topics were considered. It is
demonstrated that the spectral dimension of the spacetime where
Quantum Einstein Gravity lives in is equal to 2 microscopically
while to 4 on macroscopic scales [2]. It is also shown that the
world with a quantum group symmetry has a scale-dependent fractal
dimension at short scales to describe a phenomenon appeared in the
quantum gravity [3]. A kind of field theory which is Lorentz
invariant, power-counting renormalizable, ultraviolet finite and
causal is proposed to search for a consistent theory of quantum
gravity [4, 5].

More than 80 years ago Kaluza and Klein put forward the model that
our universe has more than four dimensions [6, 7]. The extra
spatial dimensions belonging to the modern Kaluza-Klein theory are
chosen to be compact and small to unify all interactions in
nature. Their characteristic size is of the order of the Planck
length. In addition the quantum gravity such as string theory or
brane-world scenario is developed to reconcile the quantum
mechanics and gravity with the help of introducing the additional
spatial dimensions. It seems to be reasonable to describe the
extra space by means of fractal geometry because of the quantum
fluctuations on the Planck scale space. The Kaluza-Klein model
with non-integer number of additional dimensions needs to be
developed in particular in various directions. The Kaluza-Klein
theory involving a fractal extra dimension $D$ within the range
$0<D<2$ was put forward in Ref. [8].

A resemble macroscopic quantum effect describing the attractive
force between two conducting and neutral parallel plates was put
forward by Casimir more than 60 years ago [9-12]. The Casimir
effect appears due to the disturbance of the vacuum of the
electromagnetic field induced by the presence of boundary. The
Casimir effect becomes a powerful tool to research on the model of
Universe with more than four dimensions because the effect has
something to do with the additional dimensions and the precision
of the effect measurement has been greatly improved practically
[13-16] and the effect is an observable and trustworthy
consequence of the existence of quantum fluctuations. More efforts
have been made to the Casimir effect for parallel plates within
the frame of Kaluza-Klein model with the integer number of
additional dimensions [17-29]. It has been shown that the
extra-dimension influence was manifest and distinct. It is
significant and practical to study the Casimir effect for parallel
plates in the spacetimes involving fractal extra compactified
dimensions. We wonder how the influence from fractal extra
dimensions on the Casimir effect for the parallel-plate system.
This problem, to our knowledge, has not been considered. As a
starting point and for simplicity, we choose that the background
has one fractal additional compactified dimension. The main
purpose of this letter is to explore the Casimir effect for two
parallel plates in the spacetime with one fractal extra
compactified spatial dimension. We obtain the expressions of the
total vacuum energy and force respectively. Further we obtain the
Casimir energy and Casimir force by means of the zeta-function
regularization [30, 31] that is used to leave finite results via
throwing the divergent terms and our results are not
regularization dependent. During the process we will perform the
regularization of multiple zeta function with one arbitrary
exponent again. We will focus on the influence from fractal
dimensionality of the additional space on the Casimir effect for
two parallel plates. The fractal dimensionality is within the
limit between $0$ and $2$ instead of being exact like an integer.
We hope to constrain that. Our discussions and conclusions are
listed in the end.

The content of the fractal extra compactified dimension in the
Kaluza-Klein theory were introduced in Ref. [8]. According to
fractal geometry the scale dimension $\delta=D-D_{T}$ is defined
as,

\begin{equation}
\delta\equiv\frac{d(\ln L)}{d(\ln\frac{l}{\lambda})}
\end{equation}

\noindent where $L$ is the main fractal variable denoting a length
of a fractal curve, an area of a fractal surface, etc.. The
coefficient $l$ is the measurement scale. The coefficient
$\lambda$ is the resolution of the measurement. Here $D$ is the
fractal dimension. $D_{T}$ is the topological dimension and
$D_{T}=1$ for a curve, $D_{T}=2$ for a surface. According to Eq.
(1) that $\delta$ is a constant leads,

\begin{equation}
L=L_{0}(\frac{l}{\lambda})^{\delta}
\end{equation}

\noindent where the length $L_{0}$ is measured when $\lambda=l$
and of the order of the Planck length. Within the Kaluza-Klein
issue the field must be periodic in the fifth coordinate denoted
as $x^{5}$, leading to the appearance of an infinite tower of
solutions with a quantized $x^{5}$ component of the momentum like
$q_{n}=\frac{2\pi n}{L}$ where $L$ satisfies Eq. (2). It may be
assumed that the measurement scale $\lambda$ in Eq. (2)
corresponds to $\frac{2\pi}{q_{n}}$ since any constant factor may
be absorbed in the value of $l$. The equation is obtained as
follow,

\begin{equation}
q_{n}L_{0}(\frac{lq_{n}}{2\pi})^{\delta}=2\pi n
\end{equation}

\noindent Certainly the spectrum of momentum is

\begin{equation}
q_{n}=2\pi(\frac{n}{L_{0}l^{\delta}})^{\frac{1}{1+\delta}}
\end{equation}

\noindent where $n$ is a nonnegative integer. This tower of
solutions will recover to be the tower in the regular
five-dimensional Kaluza-Klein theory if we choose $\delta=0$.

We start to consider the massless scalar field in the
two-parallel-plate system in the spacetime with a non-integer
numbered extra compactified dimension. The field obeys the
Dirichlet condition, leading the wave vector in the directions
restricted by the plates to be $k_{n}=\frac{n\pi}{R}$, $n$ a
positive integer and $R$ the separation of the two plates. In the
case of a fractal additional compactified dimension we find the
frequency of the vacuum fluctuation within a region confined by
two parallel plates whose separation is $R$ to be,

\begin{equation}
\omega_{Nn}=\sqrt{k^{2}+\frac{N^{2}\pi^{2}}{R^{2}}
+\frac{(2\pi)^{2}}{(L_{0}l^{\delta})^{\frac{2}{1+\delta}}}
n^{\frac{2}{1+\delta}}}
\end{equation}

\noindent where

\begin{equation}
k^{2}=k_{1}^{2}+k_{2}^{2}
\end{equation}

\noindent Here we employ the momentum of the fifth coordinate
under the fractal geometry presented in Eq. (4). $k_{1}$ and
$k_{2}$ are wave vectors in directions of the unbound space
coordinates parallel to the plates surface. Now $n$ is a
nonnegative integer. According to Ref. [10-12, 30, 31] the total
energy density of the fields in the interior of two-parallel-plate
device reads,

\begin{eqnarray}
\varepsilon(R, L_{0}, \delta)=\int
\frac{d^{2}k}{(2\pi)^2}\sum_{N=1}^{\infty}
\sum_{n=0}^{\infty}\frac{1}{2}\omega_{Nn}\nonumber\hspace{4.5cm}\\
=\frac{\pi^{2}}{4R^{3}}\frac{\Gamma(-\frac{3}{2})\zeta(-3)}{\Gamma(-\frac{1}{2})}
+\frac{1}{4\pi}\frac{\Gamma(-\frac{3}{2})}{\Gamma(-\frac{1}{2})}
M_{2}(-\frac{3}{2}; \frac{\pi^{2}}{R^{2}},
\frac{(2\pi)^2}{(L_{0}l^{\delta})^{\frac{2}{1+\delta}}}; 2,
\frac{2}{1+\delta})
\end{eqnarray}

\noindent in terms of the multiple zeta function with one
arbitrary exponent. The general multiple zeta function with
arbitrary exponents $M_{2}(s; a_{1}, a_{2}; \alpha_{1},
\alpha_{2})$ is defined as,

\begin{equation}
M_{2}(s; a_{1}, a_{2}; \alpha_{1},
\alpha_{2})=\sum_{n_{1},n_{2}=1}^{\infty}(a_{1}n_{1}^{\alpha_{1}}
+a_{2}n_{2}^{\alpha_{2}})^{-s}
\end{equation}

\noindent This kind of zeta function has been regularized with
neglecting the small contribution in Ref. [30, 31]. Here we
utilize the standard method to regularize the multiple zeta
function with one arbitrary exponent $M_{2}(s; a_{1}, a_{2}; 2,
\alpha_{2})$ which will be used in this work as follow,

\begin{eqnarray}
M_{2}(s; a_{1}, a_{2}; 2, \alpha_{2})\nonumber\hspace{7cm}\\
=\sum_{n_{1},n_{2}=1}^{\infty}(a_{1}n_{1}^{2}
+a_{2}n_{2}^{\alpha_{2}})^{-s}\nonumber\hspace{6cm}\\
=-\frac{1}{2}a_{2}^{-s}\zeta(\alpha_{2}s)+\frac{1}{2}
a_{2}^{-s}\sqrt{\frac{\pi
a_{2}}{a_{1}}}\frac{\Gamma(s-\frac{1}{2})}{\Gamma(s)}
\zeta(\alpha_{2}(s-\frac{1}{2}))\nonumber\hspace{1.7cm}\\
+2\pi^{s}a_{1}^{-\frac{s}{2}-\frac{1}{4}}a_{2}^{-\frac{s}{2}+\frac{1}{4}}
\sum_{n_{1},n_{2}=1}^{\infty}(\frac{n_{1}}{n_{2}^{\frac{\alpha_{2}}{2}}})^{s-\frac{1}{2}}
K_{-(s-\frac{1}{2})}(2\pi\sqrt{\frac{a_{2}}{a_{1}}}n_{1}n_{2}^{\frac{\alpha_{2}}{2}})
\end{eqnarray}

\noindent where $K_{\nu}(z)$ is the modified Bessel function of
the second kind and drops exponentially with $z$. We keep the
small contribution denoted as the modified Bessel function term.
We make use of the regularization of the multiple zeta function
with one arbitrary exponent in Eq. (9) to obtain the Casimir
energy per unit area of a system consisting of two parallel plates
in the spacetime with one fractal extra compactified spatial
dimension,

\begin{eqnarray}
\varepsilon_{C}=-\frac{\pi^{2}}{720}\frac{1}{R^{3}}
+\frac{1}{2\pi^{\frac{3}{2}}}\frac{1}{(L_{0}l^{\delta})^{\frac{3}{1+\delta}}}
\Gamma(-\frac{3}{2})\zeta(-\frac{3}{1+\delta})
-\pi^{2}\frac{R}{(L_{0}l^{\delta})^{\frac{4}{1+\delta}}}
\Gamma(-2)\zeta(-\frac{4}{1+\delta})\nonumber\\
-\frac{1}{R(L_{0}l^{\delta})^{\frac{2}{1+\delta}}}
\sum_{n_{1},n_{2}=1}^{\infty}(\frac{n_{2}^{\frac{1}{1+\delta}}}{n_{1}})^{2}
K_{2}(4\pi\frac{R}{(L_{0}l^{\delta})^{\frac{1}{1+\delta}}}
n_{1}n_{2}^{\frac{1}{1+\delta}})\hspace{3cm}
\end{eqnarray}

\noindent It is clear that the $\Gamma(-2)$ term is equal to
infinity. According to the procedure in Ref. [32] we can subtract
the part without plates to renormalize the Casimir energy density
because the subtracted part compensates exactly the divergent term
involving $\Gamma(-2)$. In the absence of plates the vacuum energy
density is
$\varepsilon_{0}=\int\frac{d^{3}\kappa}{(2\pi)^{3}}\sum_{n=0}^{\infty}
\frac{1}{2}\sqrt{\kappa^{2}+\frac{(2\pi)^{2}}{(L_{0}l^{\delta})^{\frac{2}{1+\delta}}}
n^{\frac{2}{1+\delta}}}=\pi^{2}\Gamma(-2)\zeta(-\frac{4}{1+\delta})
\frac{1}{(L_{0}l^{\delta})^{\frac{4}{(1+\delta)}}}$. The
expression for the renormalized Casimir energy density becomes,

\begin{eqnarray}
\varepsilon_{C}^{ren}=\varepsilon_{C}-R\varepsilon_{0}\nonumber\hspace{9cm}\\
=-\frac{\pi^{2}}{720}\frac{1}{R^{3}}
-\frac{\pi^{\frac{\delta-2}{1+\delta}}}{2(1+\delta)}
\frac{\Gamma(\frac{3}{2(1+\delta)})\Gamma(\frac{4+\delta}{2(1+\delta)})}
{\Gamma(\frac{3}{2})}\zeta(\frac{4+\delta}{1+\delta})
(\sin\frac{3\pi}{2(1+\delta)})\frac{1}{(L_{0}l^{\delta})^{\frac{3}
{1+\delta}}}\nonumber\\
-\frac{1}{R(L_{0}l^{\delta})^{\frac{2}{1+\delta}}}
\sum_{n_{1},n_{2}=1}^{\infty}(\frac{n_{2}^{\frac{1}{1+\delta}}}{n_{1}})^{2}
K_{2}(4\pi\frac{R}{(L_{0}l^{\delta})}n_{1}n_{2}^{\frac{1}{1+\delta}})\hspace{3cm}
\end{eqnarray}

\noindent If the plate distance is extremely large, the
renormalized Casimir energy density reduces to be

\begin{equation}
\lim_{R\longrightarrow\infty}\varepsilon_{C}^{ren}=
-\frac{\pi^{\frac{\delta-2}{1+\delta}}}{2(1+\delta)}
\frac{\Gamma(\frac{3}{2(1+\delta)})\Gamma(\frac{4+\delta}{2(1+\delta)})}
{\Gamma(\frac{3}{2})}\zeta(\frac{4+\delta}{1+\delta})
(\sin\frac{3\pi}{2(1+\delta)})\frac{1}{(L_{0}l^{\delta})^{\frac{3}
{1+\delta}}}
\end{equation}

\noindent In order to keep the negative nature of the Casimir
energy, it is necessary to choose that
$\sin\frac{3\pi}{2(1+\delta)}>0$ corresponding to

\begin{equation}
\delta>\frac{1}{2}
\end{equation}

The number of the fractal additional dimension varies between $0$
and $2$, or equivalently the scale dimension is within the range
$-1<\delta<1$. Now we give rise to a constrain on the parameter to
narrow the region like $\delta\in(\frac{1}{2},1)$.

By means of the differentiation rule for the modified Bessel
functions $\frac{\partial}{\partial
z}K_{\nu}(z)=-\frac{1}{2}[K_{\nu-1}(z)+K_{\nu+1}(z)]$ we take the
derivative of the renormalized Casimir energy density with respect
to the plates distance to obtain the Casimir force per unit plate
area between them as,

\begin{eqnarray}
f_{C}=-\frac{\partial\varepsilon_{C}^{ren}}{\partial
R}\nonumber\hspace{3cm}\\
=-\frac{\pi^{2}}{240}\frac{1}{\mu^{4}}\frac{1}{(L_{0}l^{\delta})^{\frac{4}{1+\delta}}}
+C(\mu,\delta)
\end{eqnarray}

\noindent where the correction function $C(\mu,\delta)$ is
expressed as,

\begin{eqnarray}
C(\mu,\delta)=-\{\frac{1}{\mu^{2}}
\sum_{n_{1},n_{2}=1}^{\infty}(\frac{n_{2}^{\frac{1}{1+\delta}}}{n_{1}})^{2}
K_{2}(4\pi\mu n_{1}n_{2}^{\frac{1}{1+\delta}})\nonumber\hspace{5cm}\\
-\frac{2\pi}{\mu}\sum_{n_{1},n_{2}=1}^{\infty}
\frac{n_{2}^{\frac{3}{1+\delta}}}{n_{1}} [K_{1}(4\pi\mu
n_{1}n_{2}^{\frac{1}{1+\delta}})+K_{3}(4\pi\mu
n_{1}n_{2}^{\frac{1}{1+\delta}})]\}\frac{1}{(L_{0}l^{\delta})^{\frac{4}{1+\delta}}}
\end{eqnarray}

\noindent and

\begin{equation}
\mu=\frac{R}{(L_{0}l^{\delta})^{\frac{1}{1+\delta}}}
\end{equation}

\noindent It should be emphasized that the first term named as
$f_{0}=-\frac{\pi^{2}}{240}\frac{1}{\mu^{4}}\frac{1}{(L_{0}l^{\delta})^{\frac{4}{1+\delta}}}$
in Eq. (14) is the same as Casimir pressure on the parallel plates
involving massless scalar fields satisfying the Dirichlet
conditions in the four-dimensional spacetimes. It should be
pointed out that the correction function represents the deviation
from the fractal additional dimension. When the plates gap is
larger enough than the size of the extra dimension, the correction
function approaches to the zero,

\begin{equation}
\lim_{\mu\longrightarrow\infty}C(\mu,\delta)=0
\end{equation}

\noindent no matter what the value of $\delta$ is equal to. For a
definite value of $\delta$ the behaviour of the correction
function on the variable $\mu$ is plotted in Fig. 1. The shapes of
the function for various values of scale dimension $\delta$ within
the range as Eq. (13) respectively are similar, but they are not
superposition. In order to show how great the influence of fractal
extra dimension modifies the original Casimir force between the
parallel plates we introduce an ratio as a function of $\mu$ and
$\delta$,

\begin{equation}
q(\mu,\delta)=\frac{C(\mu,\delta)}{f_{0}}
\end{equation}

\noindent We show the function $q(\mu,\delta)$ for a definite
value of $\delta$ in Fig. 2. It is manifest that the ratio
$q(\mu,\delta)$ is a decreasing function of $\mu$. In Fig. 3 we
demonstrate the curves of the ratio as the functions of $\mu$ for
some different values of scale dimension $\delta$. We discover
that the value of the ratio becomes greater wholly when the scale
dimension is larger, which means that the greater scale dimension
will bring about stronger influence from the fractal additional
dimension. Having compared our recent results with those in Ref.
[23-25] we find that the magnitude of the Casimir force in the
case of integer number of additional dimension and in the fractal
one are different. Within the scope that measurement reaches the
enough precision, we can distinguish whether the revisions are
from the extra compactified space with integer dimensionality or
fractal one.

In this letter the Casimir effect for parallel plates in the
spacetime with one fractal extra compactified dimension is
discussed. We obtain the total vacuum energy density according to
the Kaluza-Klein with one non-integer dimension [8]. We perform
the regularization of multiple zeta function with one arbitrary
exponent in order to regularize the total vacuum energy density to
obtain the Casimir energy density. It is found that the sign of
the renormalized Casimir energy remains negative if the scale
dimension $\delta>\frac{1}{2}$. The renormalized Casimir energy
density is the difference between the Casimir energy density
within two parallel plates and the one in the absence of plates.
The expression of Casimir force per unit area on the plates is
also obtained. Certainly the Casimir force has something to do
with the extra spatial dimensions. The larger scale dimension
leads the greater modification on the Casimir force. As the
separation of the two parallel plates is larger enough than the
size of fractal extra dimension, the fractal-extra-dimension
influence on the Casimir force will be weaker. The curves of the
Casimir force for parallel plates in the presence of additional
compactified space with integer-numbered dimension or
fractal-numbered dimension are not superposition. They are
distinct. We open up a new direction to detect whether the
dimensionality of extra compactified space is integer or fractal
if the experimental device reaches to be sufficiently sensitive.
Our research can be generalized to the case that the background
has multiple fractal additional compactified spatial dimension.

\vspace{3cm}

\noindent\textbf{Acknowledgement}

This work is supported by NSFC No. 10875043 and is partly
supported by the Shanghai Research Foundation No. 07dz22020.

\newpage
\begin{figure}
\setlength{\belowcaptionskip}{10pt} \centering
  \includegraphics[width=15cm]{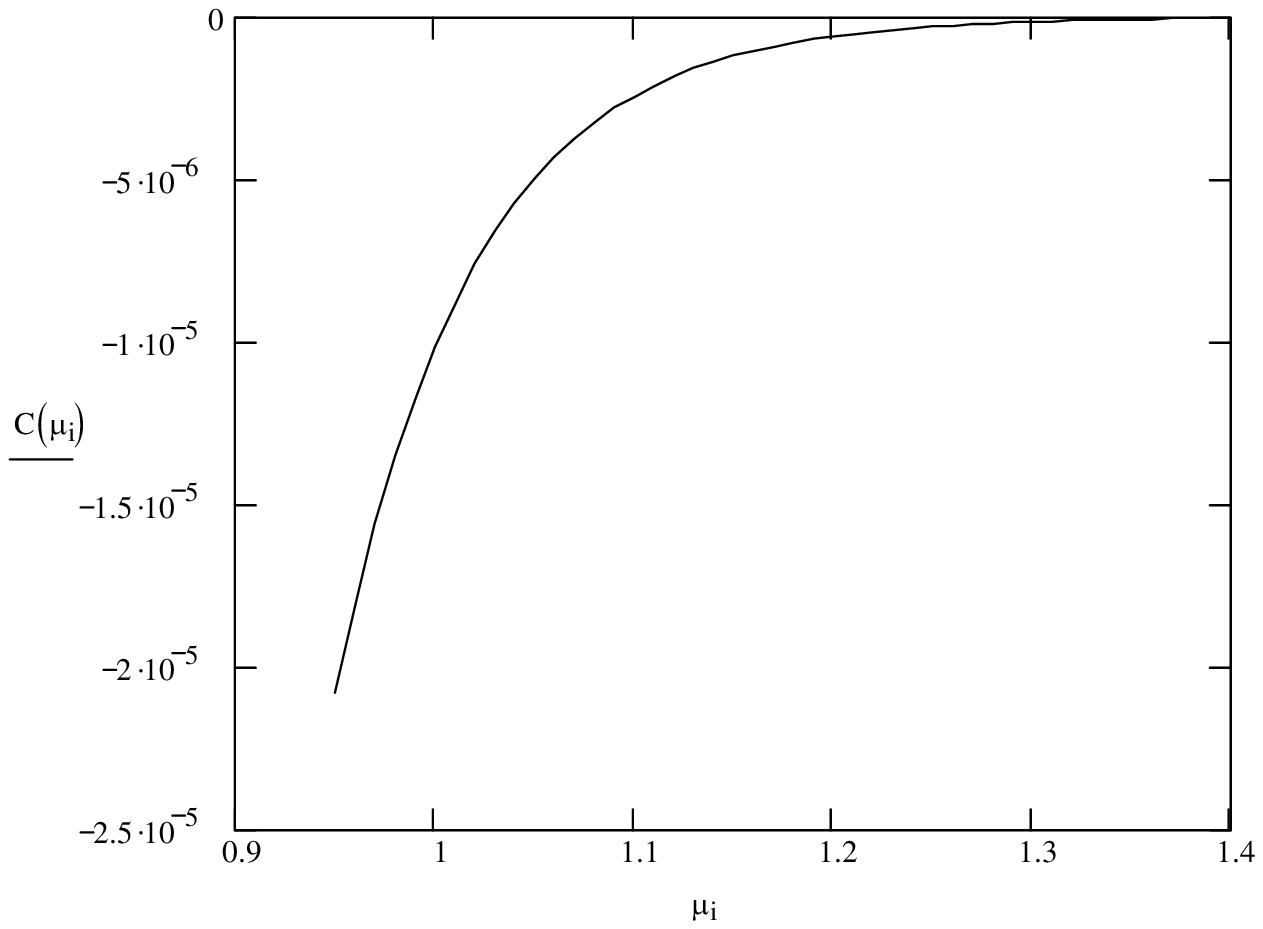}
  \caption{The curve of the correction function of $\mu=\frac{R}{(L_{0}l^{\delta})^{\frac{1}{1+\delta}}}$
  in the presence of one fractal extra compactified dimension for $\delta=0.6$.}
\end{figure}

\newpage
\begin{figure}
\setlength{\belowcaptionskip}{10pt} \centering
  \includegraphics[width=15cm]{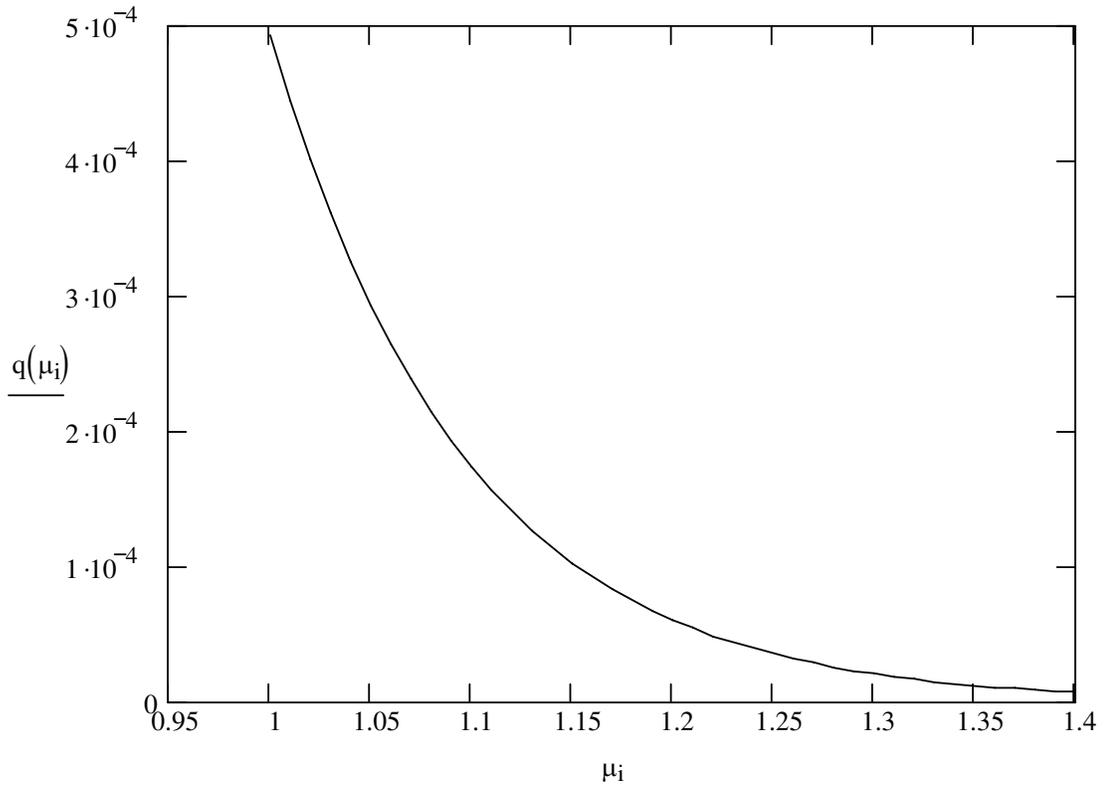}
  \caption{The behaviour of the ratio $q(\mu, \delta)=\frac{C(\mu, \delta)}{f_{0}}$
  as a function of $\mu=\frac{R}{(L_{0}l^{\delta})^{\frac{1}{1+\delta}}}$ in
  the presence of one fractal extra compactified dimension for $\delta=0.6$.}
\end{figure}

\newpage
\begin{figure}
\setlength{\belowcaptionskip}{10pt} \centering
  \includegraphics[width=15cm]{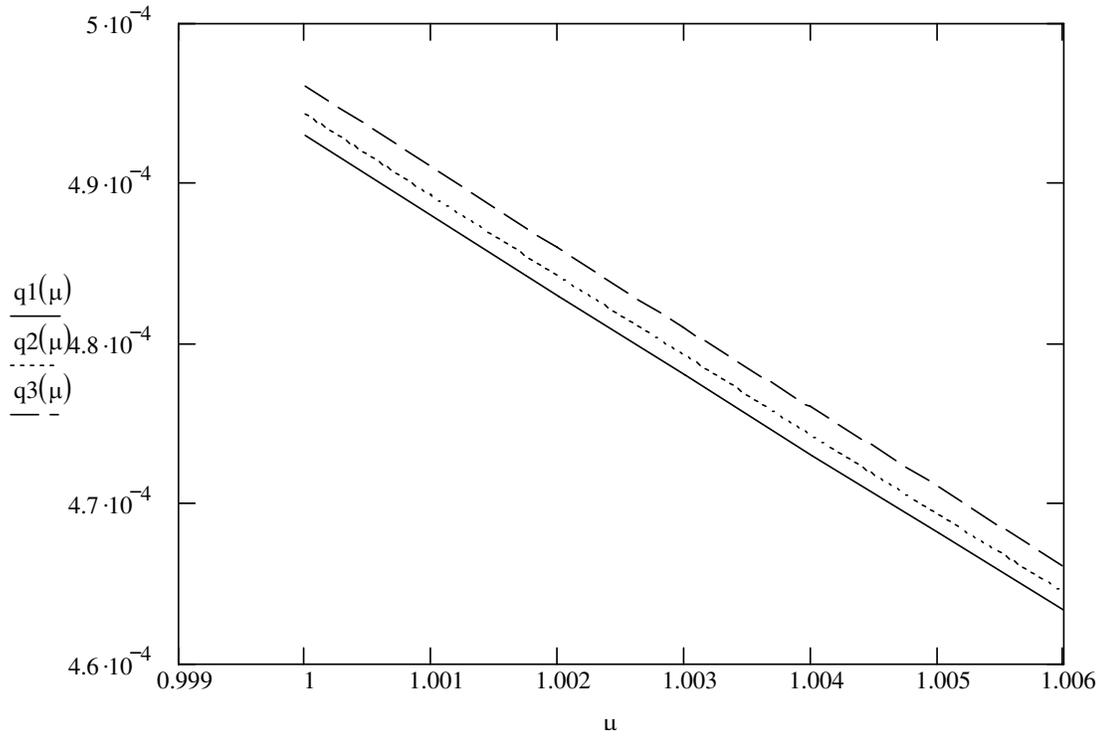}
  \caption{The solid, dot, dashed curves of the ratio $q(\mu, \delta)=\frac{C(\mu, \delta)}{f_{0}}$
  as a function of $\mu=\frac{R}{(L_{0}l^{\delta})^{\frac{1}{1+\delta}}}$ in
  the presence of one fractal extra compactified dimension for $\delta=0.6, 0.75, 0.9$
  respectively.}
\end{figure}

\end{document}